\documentclass[a4paper,fleqn,usenatbib,useAMS]{mnras}
\usepackage{newtxtext,newtxmath}
\usepackage{ae,aecompl}
\usepackage[T1]{fontenc}

\usepackage{graphicx}
\usepackage{amsmath}
\usepackage{subfig}
\usepackage{multirow}
\usepackage{caption,booktabs}

\title{Insights into physical conditions and magnetic fields from high redshift quasars}

\author[B. Lee et al.]{
Bomee Lee$^{1,2}$\thanks{E-mail:\href{mailto:bomee@kasi.re.kr}{bomee@kasi.re.kr}}
{Ranga-Ram Chary$^{1}$}
\\
$^{1}$MS314-6, Infrared Processing and Analysis Center, California Institute of Technology, Pasadena, CA 91125, USA\\
$^{2}$Korea Astronomy and Space Science Institute, 776 Daedeokdae-ro, Yuseong-gu, Daejeon 34055, Korea
}

\date{Accepted XXX. Received YYY; in original form ZZZ}

\pubyear{2022}

\begin{document}
\label{firstpage}
\pagerange{\pageref{firstpage}--\pageref{lastpage}}
\maketitle

\begin{abstract}

We use archival WISE and Spitzer photometry to derive optical line fluxes for a sample of distant quasars at z$\sim$5.9. We find evidence for exceptionally high equivalent width [OIII] emission (rest-frame EW$\sim$400\AA) similar to that inferred for star-forming galaxies at similar redshifts. The median H$\alpha$ and H$\beta$ equivalent widths are derived to be $\sim$400\AA\ and $\sim$100\AA\ respectively, and are consistent with values seen among quasars in the local Universe, and at $z\sim2$. After accounting for the contribution of photoionization in the broad line regions of quasars, we suggest that the [OIII] emission corresponds to strong, narrow-line emission likely arising from feedback due to massive star-formation in the quasar host. The high [OIII]/H$\beta$ line ratios
can uniquely be interpreted with radiative shock models, and translate to magnetic field strengths of $\sim$8 $\mu$Gauss with shock velocities of $\sim$400\,km/s. Our measurement implies that strong, coherent magnetic fields were present in the interstellar medium at a time when the universe was $\lesssim1$ billion years old. Comparing our estimated magnetic field strengths with models for the evolution of galaxy-scale fields, favors high seed field strengths exceeding 0.1 $\mu$Gauss, the first observational constraint on such fields. This high value favors scenarios where seed magnetic fields were produced by turbulence in the early stages of galaxy formation.
Forthcoming mid-infrared spectroscopy with the James Webb Space Telescope will help constrain the physical conditions in quasar hosts further.

\end{abstract}

\begin{keywords}
editorials, notices -- miscellaneous
\end{keywords}

\section{Introduction}

It is well known that magnetic fields permeate the interstellar medium of galaxies. The origin of these galaxy-scale
micro-Gauss magnetic fields is however unclear \citep[e.g.][for a review]{kul_zwei2008}. One idea is that primordial seed fields \citep{ratra1992} which are 
a relic of the inflation era, get amplified during structure formation. These seed fields which are estimated to be of order 10$^{-17}-10^{-20}$ Gauss get amplified
by up to ten orders of magnitude through the alpha-omega mean-field dynamo \citep{Ruzmaikin1985}.
Observational measurements on the amplitude of the seed field, and whether it even exists, are unclear although the angular power spectrum of the cosmic microwave background places
some upper limits at the few nanoGauss (nG) level \citep{Planck2016}, several orders of magnitude above expectations. 

An alternative to primordial seed fields posits that turbulence in the interstellar medium (ISM) during structure
formation can result in the formation of current eddies which produce a protogalactic magnetic field with amplitudes of order 0.01--1 $\mu$Gauss ($\mu$G)  \citep{Biermann1951, Ruzmaikin1988, kul97}. This field can get amplified as baryons collapse into dark matter halos by an amplitude that depends on the dynamical timescale of the ISM and the turbulence power,
with subsequent partial dissipation through ambipolar diffusion and magnetic reconnection \citep{kul97,how97}.

By measuring magnetic field strengths in galaxies at early cosmic times, we can distinguish between these two scenarios. Attempts to do so have been undertaken through radio polarization \citep{Mao2017}, through Zeeman splitting of radio lines \citep{Wolfe2008} and through optical spectroscopy \citep{Bernet2008}. However, these are all at late cosmic times,
when the Universe was $>3$ Gyr old, when the number of dynamical timescales is very large, thus diluting the constraints on the
original seed field. Until now, the capabilities to undertake these measurements at cosmic times of $<$1 Gyr have been missing.  

Emission lines from ionized gas, particularly H$\alpha$, H$\beta$ and [\ion{O}{3}] are a tracer of ISM conditions, particularly the magnetic field strength, the shock velocity, gas density and metallicity \citep{all08,kew13}. 
At high redshifts, these diagnostic lines are redshifted to mid-infrared wavelengths. Although this will be remedied imminently through spectroscopic observations with the {\it James Webb Space Telescope}, it has not been possible thus far, to place constraints on the line strengths. In the case of star-forming, emission-line galaxies at high redshift, a pioneering approach which utilized the excess emission in broad, mid-infrared bandpasses yielded strong evidence for high equivalent width nebular emission but not the line ratio accuracies needed to characterize the interstellar medium \citep[e.g][]{Chary2005, Shim2011, Faisst2016}. 

In comparison, quasars, by being the most luminous sources in the distant Universe, have multi-wavelength photometry that has much higher signal-to-noise ratio than for most star-forming galaxies.
These are from a combination of ground- and space-based surveys. Here, we leverage this high quality photometry to derive rest-frame optical emission-line properties of a sample of $z\sim6$ quasars. We compare these derived properties of the high redshift quasars which have luminosities exceeding $10^{46}$ erg\,s$^{-1}$ with the most luminous active galactic nuclei (AGN) in the local Universe, and a $z\sim2$ sample studied with VLT/SINFONI \citep{Vietri}. By fitting radiative shock models to their optical emission line ratios we are able
to derive magnetic field strengths and shock velocities of both the high redshift and local AGN. We present our results in the context of the expected evolution of galaxy scale magnetic 
fields and use our new constraints to distinguish between different scenarios for the seeding of magnetic fields.

Throughout this paper, we adopt a (0.3,0.7,70) flat, $\Lambda$CDM cosmology.

\section{Data and Sample Selection}

Our initial sample consists of a list of 488 quasars at $z > 5$ that have been spectroscopically confirmed and published as of 2018 December 31 \citep{ros20}, and 32 additional quasars at $z >5$ published in 2019\footnote{https://github.com/d80b2t/VHzQ}. We collect multi-wavelength photometric data between the optical $y-$band ($\sim0.9\,\mu$m) and 8 $\mu m$ from archival imaging data on those $z>5$ quasars. The data includes $JHK/Ks$ photometry from WFCAM/UKIRT and VIRCAM/VISTA \citep{ros20}, $y$ band from Pan-STARRS1 DR2 \citep{cha16}, and mid-infrared photometry in the {\it Spitzer}/IRAC channels (4 channels: 3.6, 4.5, 5.8, 8.0 $\mu m$) and WISE (3.4, 4.6 $\mu m$; AllWISE) bands obtained from the SEIP (Spitzer Enhanced Imaging Products)\footnote{https://irsa.ipac.caltech.edu/}. We use flux densities in the IRAC bands measured in a 3.8\arcsec diameter aperture, and apply an aperture correction. For WISE, we use point spread function (PSF) profile-fitting photometry. 

\begin{figure}
\centering
\includegraphics[width=\linewidth]{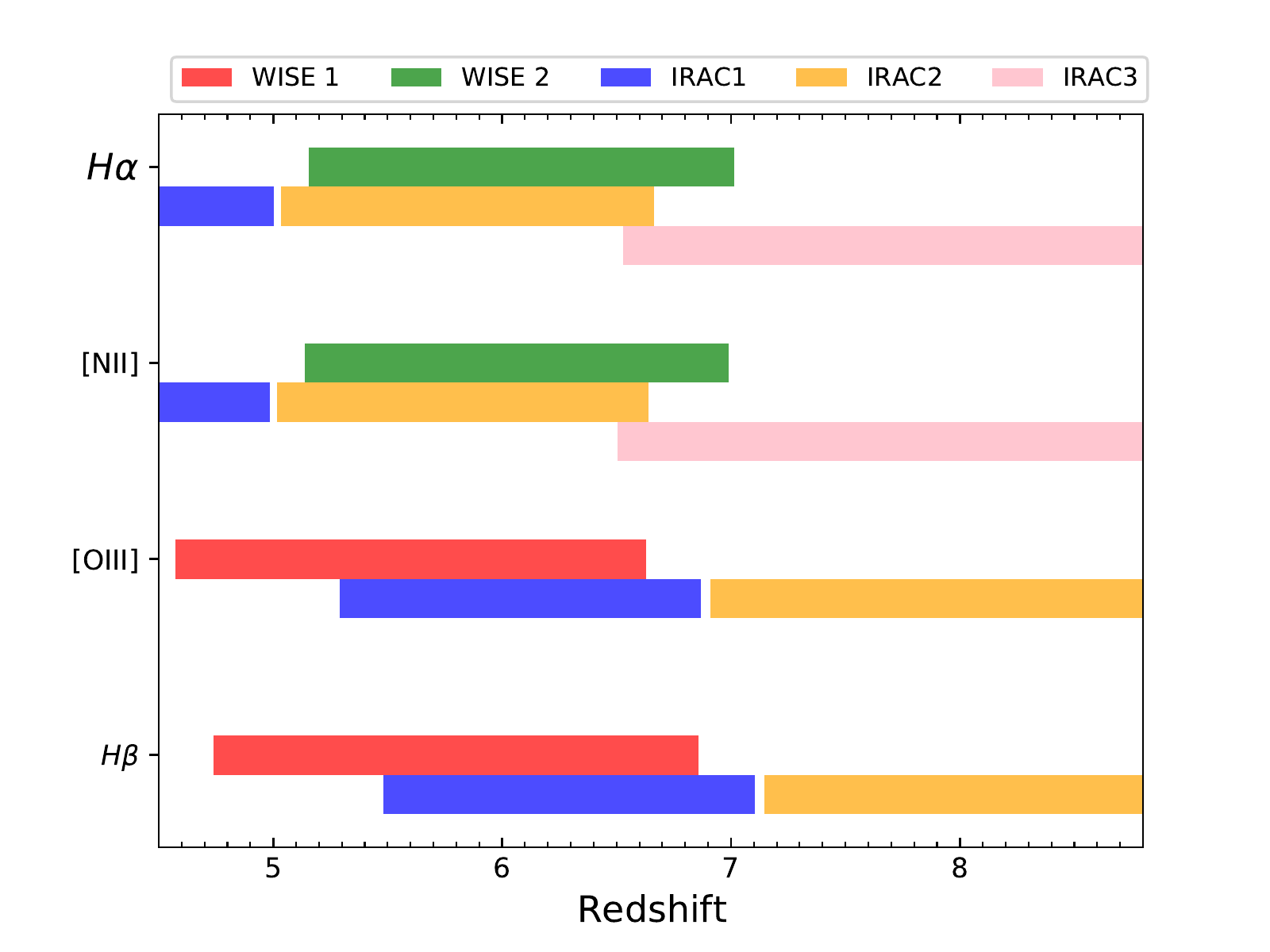}
\caption{ An illustration showing how different bands sample different rest-frame optical emission lines (H$\alpha$, [NII], [OIII], and H$\beta$) at different redshifts for $4.5<z<8.8$. IRAC 1, 2, 3 channels are shown as blue, orange, and pink bars, respectively, while WISE 3.4 and 4.6 $\mu m$ bands (WISE 1 and 2) are shown as red and green bars. For each filter, we adopt the wavelength range where the spectral response is greater than 10\% of the maximum to calculate the redshift ranges for individual emission lines. There is a gap between the spectral response curve of IRAC 1 and IRAC 2 at $5 < z < 5.03$ so that we cannot derive H$\alpha$ and [NII] line fluxes with IRAC photometry. Also, at $z < 5.3$, [OIII] and H$\beta$ are solely measured with WISE 1. }
\label{fig:filters}
\end{figure}

We apply an additional set of strict criteria on the sample, as described below:
\begin{itemize}
\item Quasars are selected to have signal-to-noise ratio (SNR) $>5$ in both IRAC 1 (3.6 $\mu m$) and 2 (4.5 $\mu m$) channels. 
\item Only quasars at $z > 5.03$ are selected. As shown in Figure~\ref{fig:filters}, there is a reduction in throughput between the 3.6 $\mu m$ and 4.5 $\mu m$ IRAC channels which implies that constraints are impossible on the H$\alpha$ emission line fluxes at redshifts of $5 < z < 5.03$.
\item Quasars at $z< 5.3$ without WISE 3.4 $\mu m$ are excluded because H$\beta$ and [\ion{O}{3}] can only be estimated from WISE 3.4 $\mu m$ as shown in Figure ~\ref{fig:filters}. 
\item NIR photometry in at least one of $J,H,K$ bands with SNR$ > 5$ is required. 
\end{itemize}

Through these cuts, we reduce the initial sample to 53 quasars at $5.03 < z < 6.3$. 

The pixel scale of Spitzer is 1.2$\arcsec$ and the full width at half maximum (FWHM) of the point spread function is $\sim$2$\arcsec$. Since quasars detected in the {\it Spitzer}/IRAC or WISE catalog can suffer from blending (i.e. source confusion) which biases the photometry, we use the Pan-STARRS1 high resolution catalogs to exclude sources where there may be blending with nearby objects. We also check the quasar images visually to ensure that they  are isolated sources. We apply corrections for Galactic extinction to all the photometry using the \citet{Schlafly} estimates but we note
that the E(B-V) values are small, with a median of about 0.02. We also applied color corrections to the mid-infrared (IRAC and WISE) photometry using the correction values provided in the IRAC and WISE instrument handbook. 

\begin{figure}
\centering
\includegraphics[width=\linewidth]{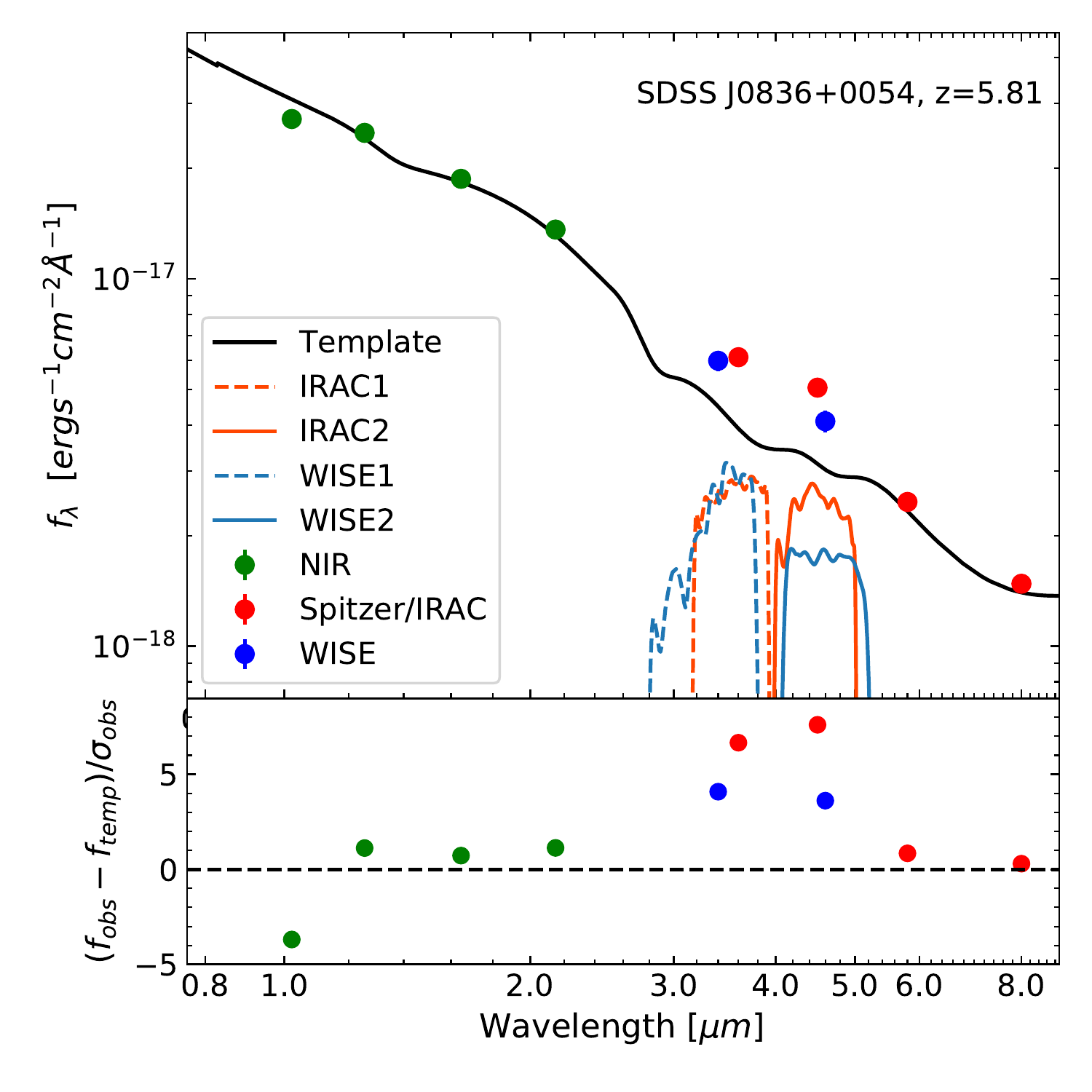}
\caption{An example fit of SDSS J0836+0054, a quasar at z$=5.81$ with its observed multi-wavelength photometry from Pan-STARRS1 (y: green), WFCAM/VISTA (JHK: green points), Spitzer (IRAC 1,2,3,4 channels: red points), and WISE (3.4, 4.6 $\mu m$: blue points). Black line represents the continuum template best-fitted to near-IR bands and IRAC 3, 4 channels. The relative transmission curves of IRAC 1,2 (orange) and WISE 1, 2 (light blue) are overplotted. The slight difference in bandpass between IRAC and WISE bands allows an estimation of rest-frame optical spectra of quasars at $z > 5$ as shown in Figure~\ref{fig:filters}. Bottom panel: The difference between observed flux and continuum flux as a function of wavelength. One can clearly see the bumps between 3 and 5 $\mu m$ due to [OIII] and H$\beta$ in the $\sim3.5\mu$m bandpasses and H$\alpha$+[NII] in the $\sim$4.5$\mu$m bandpasses.}
\label{fig:BFsed}
\end{figure}

\section{Estimates of the rest-frame optical emission line strengths}

\subsection{Fitting the Spectral Energy Distribution}

\begin{figure*}
\centering
\includegraphics[width=\linewidth]{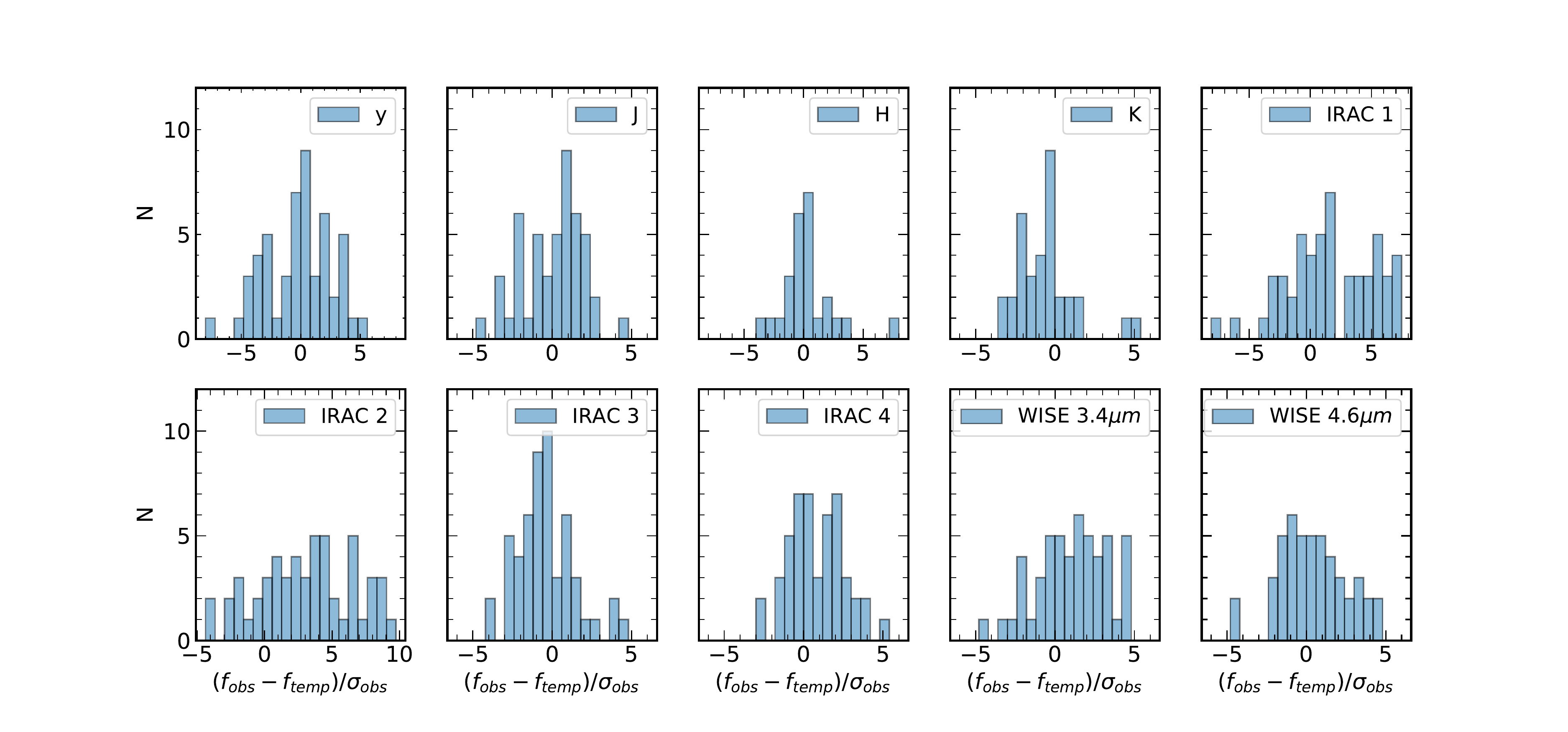}
\caption{Histograms of differences between observed flux ($f_{obs}$) and derived continuum flux ($f_{temp}$) in each band for the quasars in the sample.
The figures show $(f_{obs}-f_{temp})/\sigma_{obs}$, in
the $y, J, H, K$, IRAC 3.6-8$\mu$m bands, and WISE 3.4$\mu m$ and 4.6$\mu m$. $\sigma_{obs}$ is the observed flux density uncertainty in that band. The abscissa values are derived from the template fitting as illustrated in the bottom panel of Figure~\ref{fig:BFsed}. We find that the distribution at IRAC 1, 2 and WISE bands are skewed to positive values, indicating that those bands are indeed contaminated by strong emission lines. The distributions of yJHK are centered on zero since they do not have strong emission lines at the redshifts of the sample. This also suggests that
variability cannot account for the excess that we see in the IRAC and WISE bands; those bands have consistent photometry despite the imaging data being taken at completely different times.}
\label{fig:dflux_hist}
\end{figure*}

As shown in Figure \ref{fig:filters}, the redshifting of strong emission lines, such as H$\alpha$, [NII], H$\beta$, and [OIII], in $z > 5$ galaxies, can introduce a photometric excess in certain bands.
These lines induce a significant difference between the observed photometry  ($f_{obs}$) and the underlying continuum flux density ($f_{temp}$), especially if the equivalent width of the lines is high.  Thus, if we can fit for the underlying
continuum flux density using templates, we can obtain an estimate of the emission line fluxes using the excess emission in the different bands. We note that the
IRAC 5.8 and 8 $\mu$m bands and the WISE 3.4$\mu m$ and 4.6$\mu m$ bands, are much noisier than the IRAC 3.6 and 4.5 $\mu$m bands. So an accurate estimate of the photometric uncertainty
in each band is crucial for fitting the underlying continuum robustly and to obtain robust signal to noise estimates of the derived line fluxes.

In addition, we make the underlying assumption that variability is not important. In reality, AGN may vary since the multiwavelength data were taken from different surveys at different times by different groups and even in a single band, the data is a combination over multiple epochs of observations. However, variability should result in either an increase or decrease in the flux density in a band - if the dominant trend in the flux density residuals after fitting has a positive excess, it would imply that there is an additional contribution to the emission on top of the intrinsic variability of the source. Secondly, since the sources are unresolved, the flux densities we are measuring are a combination of the emission from the AGN broad line region, the narrow line region and the host galaxy. Although the sources are spectroscopically confirmed to be AGN, through rest-frame ultraviolet spectroscopy, it is unclear how much the integrated rest-frame optical
emission would actually vary. Nevertheless, we assess the role of variability on the measurements later in this section.

We first compile 23 AGN-dominated continuum templates obtained from local \citep{bro19} and composite AGN \citep{pol07, ass10}.
We exclude strong emission lines from the original templates and smooth the templates to remove any artificial patterns. We redshift the templates to the redshift of the quasars and then apply a correction for absorption by the intergalactic medium (IGM) using a recipe from \cite{ino14}. Before fitting templates to the photometry, systematic uncertainties  (3\% for ground-based near-IR, and 5\% for the {\it Spitzer} and {\it WISE} bands) are added in quadrature to the flux density uncertainties. We also limit the continuum fits to wavelengths where the photometric $SNR > 5$.
 
\begin{figure}
\centering
\includegraphics[width=\linewidth]{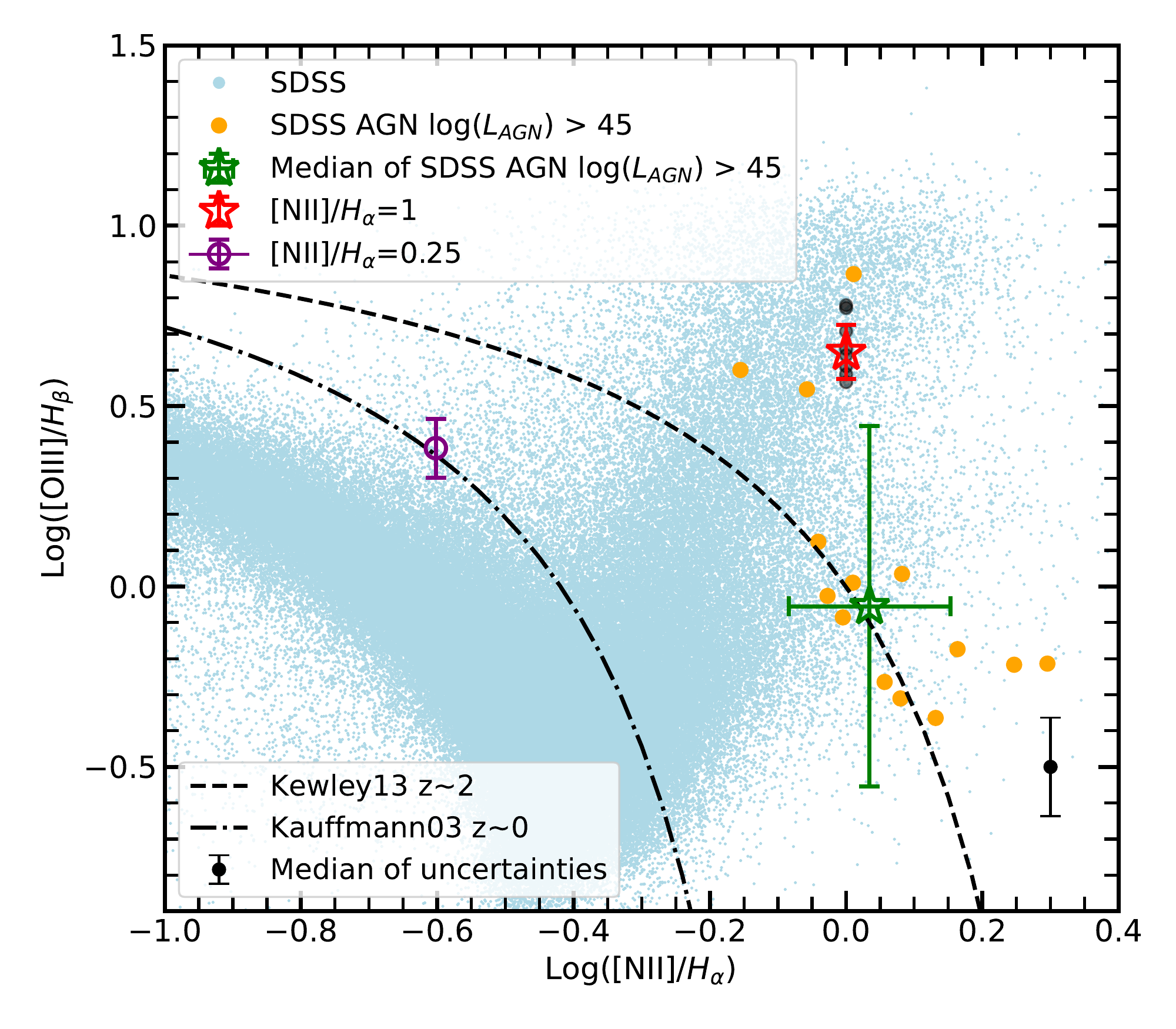}
\caption{Our constraints on the median optical emission line ratios ([OIII]/H$\beta$ vs [NII]/H$\alpha$) of quasars at z$\sim$6 are shown as the red star (with uncertainties). Black dots (which are behind the red star) represent 9 individual quasars at $5.03 < z < 6.2$ on BPT diagram. Also shown is the median ratio for luminous Sloan quasars in the local Universe (green star). The luminous Sloan AGNs with $L_{AGN} > 10^{45} erg/s$ are shown individually as the orange dots. The cyan dots are the line ratios for the complete sample of Sloan galaxies and active galactic nuclei (AGN) in the local Universe. The dot-dashed line \citep{kau03} is the demarcation at z$\sim$0 between star-forming galaxies (below the line) and AGN (above the line) while the dashed line is the same demarcation at z$\sim$2 \citep{kew13}. Black point with an error bar indicates the median uncertainty of $z\sim6$ quasars. Purple point with uncertainties is an estimate assuming $ [NII]/H\alpha$=0.25. Quasars at z$\sim$6, if they have solar metallicity, have unusually strong [OIII]/H$\beta$ which is about 1.4 $\sigma$ away from the median of local quasars. But, they are comparable to three luminous Sloan AGNs having log([OIII]/H$\beta$) $> 0.5$. }
\label{fig:bpt}
\end{figure}

As an example,  for the z=5.81 quasar shown in Figure~\ref{fig:BFsed}, H$\alpha$ and [NII] line fluxes can be estimated from the photometry in the IRAC 2 and WISE 2 bands, while H$\beta$ and [OIII] can be estimated with IRAC 1 and WISE 1 bands, as shown in Figure~\ref{fig:filters}. However, with typical quasar line widths, because H$\alpha$ is strongly contaminated with the [NII] emission line, it is difficult to separately constrain H$\alpha$ fluxes from [NII] with the broad IRAC/WISE bandpasses. Recent studies \citep{Schindler2020, deR11} have shown through rest-frame ultraviolet spectroscopy that the
metallicity of high redshift quasars seems to be very similar to those at lower redshifts, indicating that the stars in their host galaxies have already built up most of their metal abundance.  
Based on this result, we adopt [NII]/ H$\alpha$ =1 as estimated from local AGN (e.g., SDSS DR12 in Figure~\ref{fig:bpt}) and z$\sim$2.5 quasars \citep{kew13}. Since we cannot directly fit the spectra to measure line widths, we fix FWHM = 3400km/s for all lines. FWHM = 3400 km/s is chosen as a median FWHM of the MgII line in 82 quasars at $4.5 < z < 6.4$ \citep{kur07, wil10, tra11, deR11, che18} and assuming a correlation of FWHM between MgII and H$\alpha$ lines \citep{MD04, jun15, zuo15}. 

With these assumptions, we then evaluate the line fluxes as follows: 
\begin{enumerate}
\item We find the best-fit continuum template of each source by fitting to the high SNR near-IR bands ($y$, J, H, and K) and IRAC 3 and 4 channels, if available, by minimizing $\chi^2$ between observed photometry and the continuum templates convolved through the corresponding bandpass.
\item We use the excess emission at IRAC 2 and/or WISE 2, to compute H$\alpha$. We assume a Gaussian profile for the line with FWHM=3400km/s and [NII]/H$\alpha$=1. Due to the different shapes of the IRAC and WISE band transmission (as shown in Figure~\ref{fig:BFsed}), the H$\alpha$ and [NII] contribution in each band should be different, and depends on the redshift. We, therefore, calculate the fractional contribution of H$\alpha$ and [NII] in the IRAC 2 and WISE 2 bands for each object separately.

Note that the uncertainty weighted line flux values from both IRAC 2 and WISE 2 are derived, when available, for quasars at $z>5.16$ (see Figure~\ref{fig:filters}). 
\item H$\beta$ is derived by assuming an H$\alpha$/H$\beta$$=2.86$ (i.e. no dust). This is consistent with the H$\alpha$/H$\beta$ ratio of 2.6 that we find for a $z=5.37$ quasar in our sample. This object is unique since only the [OIII] line is in IRAC 1; we can therefore independently measure H$\beta$ after subtracting the [OIII] contribution from the WISE 1 excess. Although we only have this one constraint on the H$\alpha$/H$\beta$ ratio, we note that any dust in the quasar host would only increase the [OIII]/H$\beta$ ratio as we discuss later.

\item We derive [OIII] after subtracting the H$\beta$ contribution from the excess emission in IRAC 1 and/or WISE 1. Depending on the redshift, the final line flux is the weighted line flux from both the IRAC 1 and WISE 1 excess, when available, as shown in Figure~\ref{fig:filters}. At $z<5.3$, for instance, we only rely on the WISE 1 photometry to estimate [OIII] after subtracting the H$\beta$ contribution. 

\end{enumerate}
To produce robust uncertainty estimates on the fitted emission lines, we use a Monte Carlo simulation. We modify the true photometry in each of the bands by adding a random uncertainty value to the fluxes. The uncertainties are drawn from a normal Gaussian distribution with $\sigma$ corresponding to the photometric uncertainty in that band. We then repeat the fits with the synthetic photometry. The above calculation is repeated with the number of trials, $N_{tri}=600$, to obtain the $1\sigma$ uncertainty and the median of the emission line fluxes for each quasar in the sample. Note that the result does not vary significantly with larger $N_{tri}$. 
About 38\% of 53 sources have negative $(f_{obs}-f_{continum})$ at IRAC and/or WISE bands, suggesting that the emission lines are weak in these systems or that variability is dominating the fitting residuals. 
As a result, we are able to constrain H$\beta$ and [OIII] line fluxes for only a subset of 33 quasars at $5.03 < z < 6.3$. 

As mentioned earlier, since the multi-wavelength photometry of the quasars is taken at different epochs, 
it is possible that the variability of the sources between the different observing epochs can cause a significant difference between observed photometry and derived continuum flux. In general,
since the quasars were first detected in short wavelength photometry, those bands would then be skewed brighter relative to the continuum in the other bands.
To assess this, we investigate the distributions of $(f_{obs}-f_{temp})/\sigma_{obs}$  in all 10 filters, $yJHK$, IRAC 1, 2, 3, 4 channels and WISE bands in Figure~\ref{fig:dflux_hist}. 
We find that the distributions of residuals in yJHK are nearly centered on zero indicating that the residuals are mostly due to photometric noise. In contrast,
the
bands in which the expected strength of emission lines is strong (i.e. IRAC 1, 2 and WISE bands), are mostly skewed positive.
This further indicates that the contamination of the photometry due to emission lines is real and the methodology we have adopted to derive emission line constraints is robust.

In order to assess if the photometry itself may be biased due to systematics, ideally one would compare the line flux derived from photometry with spectroscopy. Unfortunately, measuring spectral lines at these redshifts is only becoming possible now with the James Webb Space Telescope. Such measurements do not exist at the present time and our work is intended to provide a prediction as to the expected line fluxes. However, we can characterize the robustness of the methodology by measuring the photometry of quasars at lower redshifts ($z\sim3$), where the line flux does not contaminate the Spitzer and WISE bands. This is discussed in the Appendix where we find that lower redshift quasars show no noticeable
excess in the IRAC bands, implying that systematics do not affect the measured photometry. 

\subsection{Optical emission line constraints at z $\sim$ 6}

In order to investigate the physical 
characteristics of z $>5$ quasars and compare them with the local sample, we plot the derived line ratios of our sample on the 
[NII]/H$\alpha$ versus [OIII]/H$\beta$ diagnostics diagram (\cite{bal81}; BPT diagram). 
As discussed earlier, we only have line flux constraints on 33 of the quasars in the sample due to large photometric uncertainties in most of the cases, and inadequacy of the templates in a small subset of cases.
In order to derive line ratios, we further need to restrict the sample to those which have high SNR line detections from our Monte-Carlo analysis.
By restricting the sample to those which have $SNR > 3$ in $H\alpha$ and $[OIII]$, we have a final sample
of 8 quasars at $5.03 < z < 6.23$ with a median redshift of $z=5.94$ with which we can study the line ratios.  In the Appendix, the catalog of 8 quasars is available in Table~\ref{table:first},~\ref{table:second}, and~\ref{table:third}. The median bolometric luminosity of this sample of quasars is $2\times10^{46}$ erg/s with a range of $3.9\times10^{45} < L_{AGN}~[erg/s] < 4.6\times10^{46}$. The median derived rest-frame equivalent widths of H$\alpha$, H$\beta$ and [OIII] are 400\AA, 100\AA\ and 440\AA. The median derived [OIII] luminosity is $2\times10^{45}$ erg/s.

Although we have to assume an [NII]/H$\alpha$ ratio as discussed earlier based on published UV spectroscopy, we can for the first time,
constrain the location of the eight $z\sim6$ quasars (black dots) on the BPT diagram (Figure~\ref{fig:bpt}). The median [OIII]/H$\beta$ ratio of the eight quasars at $5.03 < z < 6.23$ is 4.46 (red star with an uncertainty) with an assumption of solar metallicity ([NII]/H$\alpha$$=1$) and no dust. If dust in the host galaxy were present, it would extinct H$\beta$ slightly more than [OIII], and would increase the H$\alpha$/H$\beta$ ratio. 
This in turn would only serve to increase the derived contribution of [OIII] to the excess in the photometric bandpass and thereby the [OIII]/H$\beta$ ratio. We investigate this by assuming a H$\alpha$/H$\beta$ ratio of 4.5. This has been measured by \cite{jun15} in 155 luminous quasars at  $3.3 < z < 6.4$. We find that in this case, we derive a factor of 1.7 vertical increase in the [OIII]/H$\beta$ ratio.
We note that simulations suggest that the stellar masses of these objects are likely to be $>10^{10}$\,M$_{\sun}$ \citep{Marshall2020}, so the assumption of solar metallicity and high
[NII]/H$\alpha$ is well motivated.
 
If in contrast, the quasar hosts have built up their black holes faster than their underlying stellar population as is suggested by some models and observations \citep[e.g][]{deG2015, Vayner2021}, they would be expected to have
lower metallicities, and low [NII]/H$\alpha$ ratios, as well. We repeat the calculations described earlier for this unlikely scenario, which we note is inconsistent with the UV derived metallicities
of these quasars. We find that if we adopt an [NII]/H$\alpha$=0.25, the high H$\alpha$ line flux translates to a high H$\beta$ line flux as well and therefore reduces the [OIII] contribution to the photometry. This is shown as the purple point in Figure~\ref{fig:bpt}. We can therefore say with high confidence, that the [OIII]/H$\beta$ ratios of $z\sim6$ quasars are likely to be between \textbf{$2.4 - 4.5$} with a strong preference towards the high end of these values. 

For comparison, we select 14 of the most luminous local AGN (orange circles) having L$_{AGN} > 10^{45} erg/s$ with SNR$>$10 in H$\alpha$, [NII], H$\beta$, and [OIII] (hereafter, the luminous Sloan AGNs). These AGN selected from the Sloan Digital Sky Survey have been classified with the BPT diagnostic (data from SDSS DR12\footnote{http://skyserver.sdss.org/dr12/en/home.aspx}). We find that our sample of $z\sim6$ quasars, if they have solar metallicity, have unusually strong [OIII]/H$\beta$ which is above the median of local quasars (about 1.4$\sigma$). Their line ratio is comparable to the most extreme local systems as shown by the three AGNs having log([OIII]/H$\beta$) $> 0.5$. Even if their metallicities are substantially sub-solar (purple dot), the [OIII]/H$\beta$ ratio is still higher than most luminous AGN in the local Universe. Our result indicates that high-z quasars are located at the high end of the [OIII]/H$\beta$ distribution found among local AGNs. It is possible that this is due to the selection biases introduced from the high SNR requirements for detecting and characterizing lines with broadband photometry which only upcoming spectroscopic observations with JWST will be able to mitigate. 

It is worth comparing the derived equivalent widths of the high-z quasars with other samples of quasars. We find rest-frame H$\alpha$ equivalent widths of $\sim$300-800\AA\ 
at luminosities of $\sim$10$^{46}$ erg\,s$^{-1}$. This is in good agreement with the values derived for $z>5$ quasars, using a similar methodology by \citet{Leipski2014}, and similar 
to the H$\alpha$ equivalent widths of $\sim$5000 Sloan quasars \citep{Shen2011}. The difference is we extend the work of \citet{Leipski2014} to [OIII] and H$\beta$ as well. A comparison between our derived H$\beta$ equivalent widths which assumes a constant H$\beta$/H$\alpha$ ratio and those of \citet{Shen2011}, is also worthwhile. The median H$\beta$ rest-frame equivalent  width of the local SDSS quasars is 100\AA, similar to our sample,
implying that much of the excess in the bluer bandpass is because of the [OIII] doublet as discussed earlier and shown in Table \ref{table:third}.
We note that this corresponds to a substantially larger [OIII] equivalent width than that observed in $z\sim2$ quasars of similar luminosities \citep{Vietri}. 

\begin{figure*}
\centering
\subfloat{
\includegraphics[clip, width=\columnwidth]{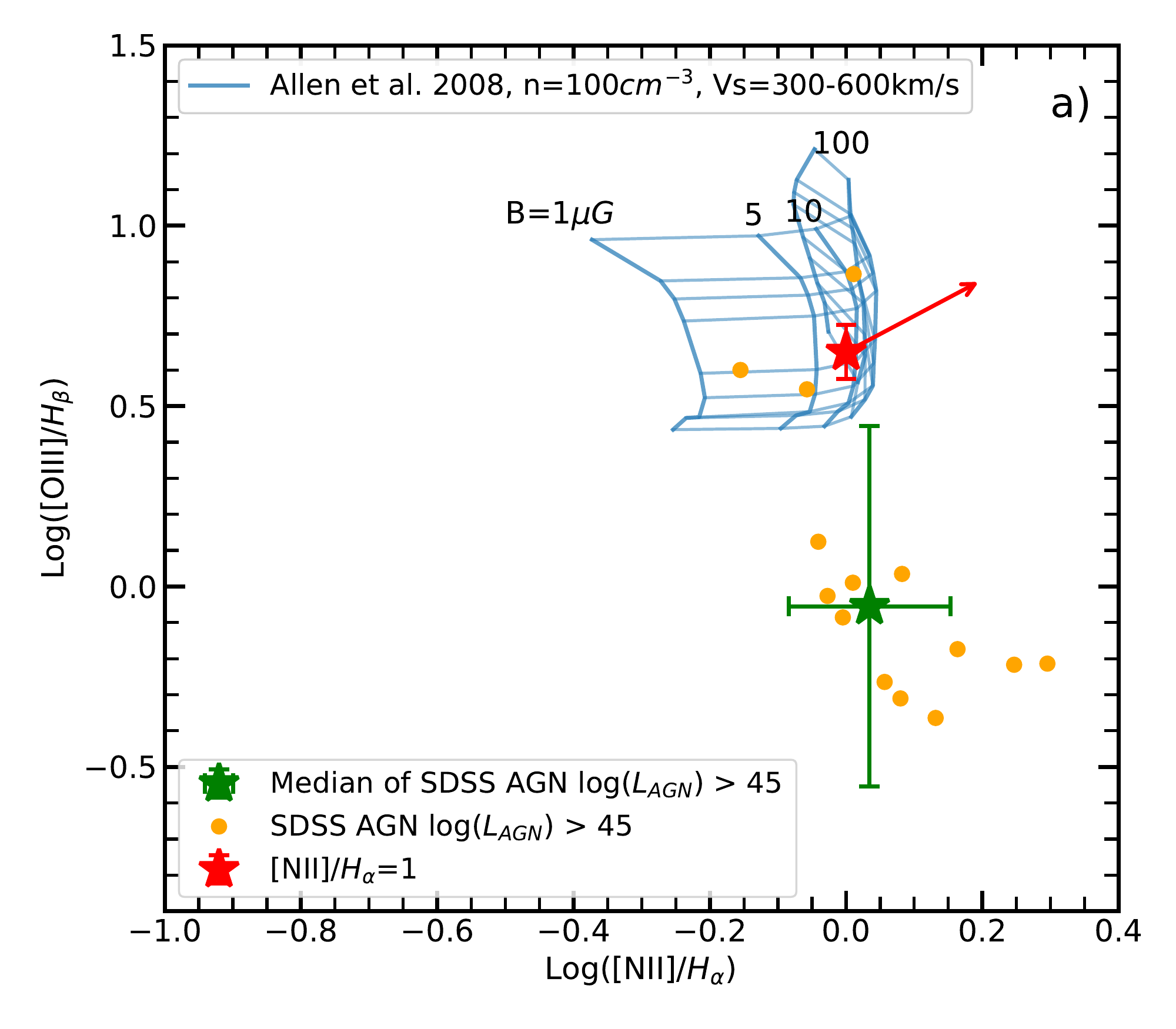}
}
\vspace{-0.3cm}
\subfloat{
\includegraphics[clip, width=\columnwidth]{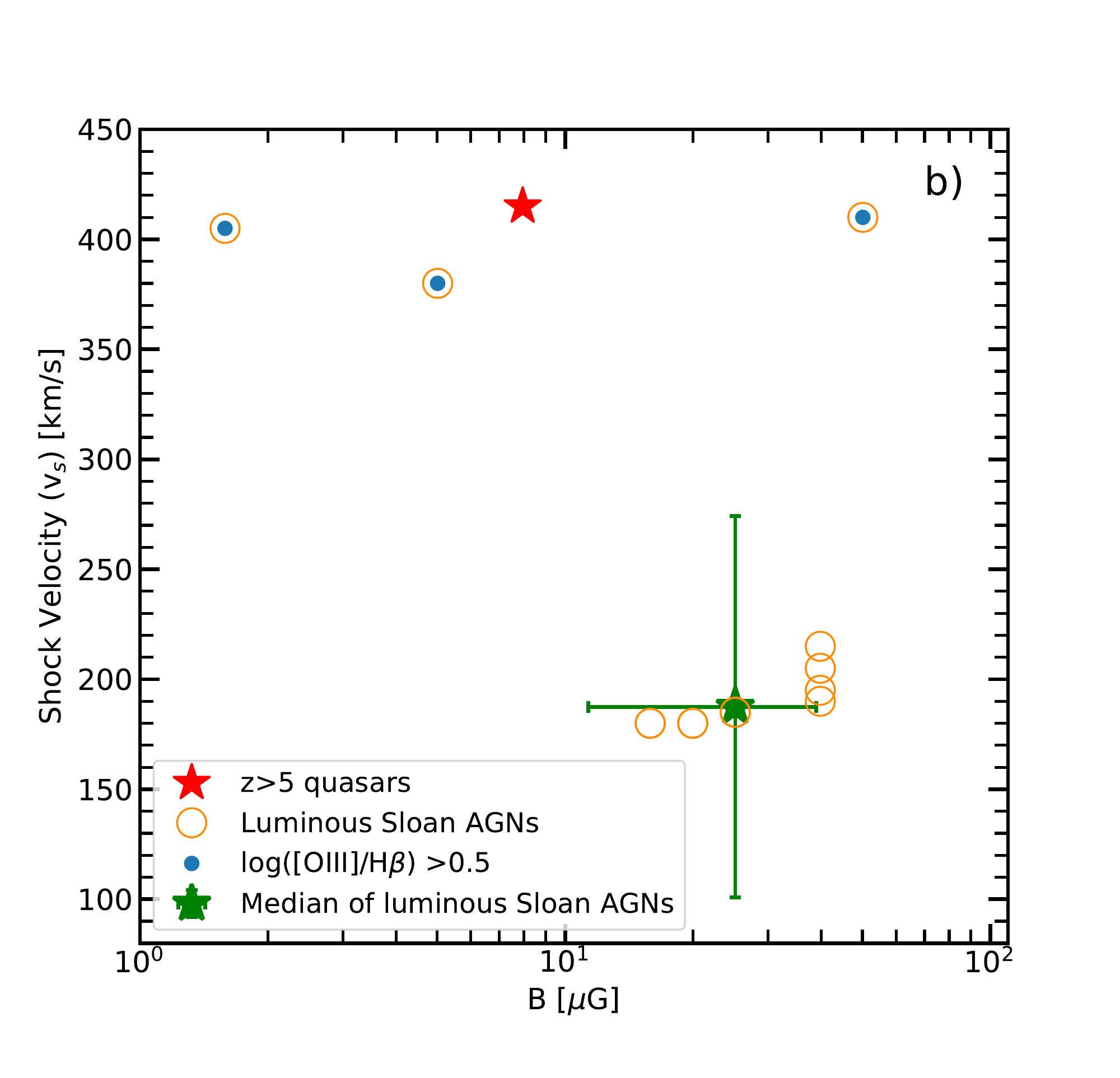}
}
\caption{Left: BPT diagram with a precursor+shock model from \protect\cite{all08}. Blue lines are precursor+shock models for a solar metallicity and n=100$cm^{-3}$  as a function of shock velocities ($V_{s}$ = 300-600$km/s$) and magnetic fields (B=1, 5, 10,  20, 50, 100 $\mu$G). The description of markers and colors are identical to Figure~\ref{fig:bpt}. Lower gas densities as expected for halo-wide averages would shift the models to the right in the same direction as the arrow. Higher metallicities would also go in the same direction. Right: The best-matched shock models ($V_{s}$ and B) of the median of $z = 5.94$ quasars (red star) and the median of luminous Sloan AGNs (green star) with its uncertainty. $V_{s}$ and B are computed with a minimization of $\chi^{2}$ between the measures of line ratios on the BPT and the precursor+shock model for a solar metallicity and $n=100cm^{-3}$ (blue lines in the left plot). We also derive $V_{s}$ and B of the individual luminous Sloan AGNs (orange circles). On average, the luminous Sloan AGNs in the local Universe have larger B-fields and lower shock velocities, while three quasars having [OIII]/H$\beta > 0.5$ (blue filled circles) have similar shock velocities with $z\sim6$. }
\label{fig:shock}
\end{figure*}

\section{Physical Implication for $z\sim6$ Quasars}

We have leveraged high quality multi-wavelength photometry to derive rest-frame optical emission line fluxes in a sample of $z\sim6$ quasars. These are the most luminous objects at early cosmic times with luminosities that are a factor of few higher than the most luminous AGN in the local Universe. Our derived [OIII]/H$\beta$ ratios are rather high compared to local AGN of comparable luminosities
but similar to the most extreme local systems. This indicates that $z\sim6$ quasars are likely to firmly lie in the Type I AGN part of the BPT diagram \citep{Stern2013}, when spectroscopic data become available.

Due to the compactness of the stellar population at high-z ($<$1 kpc) which even the next generation of telescopes will be hard-pressed to resolve against
the brightness of the quasar, the emission line fluxes that we are measuring are the sum of the narrow-line region (NLR), the broad-line region (BLR) and from stellar/ISM emission. Given the luminosity of the quasars,
the BLR is dominating the Hydrogen Balmer emission-line luminosities.
For instance, in \citet{Shen2011}, the local quasars show broad-line luminosities that are $\sim$20 times the narrow-line luminosities for H$\alpha$ and H$\beta$. Much of the broad line emission is powered by photoionization by the central source. However, the big difference is local quasar hosts are relatively quiescent with low-levels of ambient star-formation compared to a high-z galaxy sample \citep[e.g.][]{Zakamska2016}.

The measurement of the enhanced [OIII] emission is however surprising, since it likely has a significant component arising from the NLR \citep{bas05, Vietri}. In $z\sim2$ quasars of comparable bolometric luminosities as the sources studied here, the [OIII] line luminosities are$\sim$10$^{43}$ erg\,s$^{-1}$  \citep{Vietri, Kakkad2020}. The narrow component of [OIII] in those systems account for about 35\% of the total [OIII] emission. In contrast, the median [OIII] luminosity of our sample is 10$^{45.3}$\,erg\,s$^{-1}$. Since the equivalent widths of the Balmer lines are consistent between low- and high-z quasars, the implication is that
the high [OIII] luminosity of our sample arises from an additional component, likely star-formation, rather that the quasar-powered ionized outflows noted by \citet{Kakkad2020, Zakamska2016b}. 
Spatially resolved measurements of the kinematics of the [OIII] emission are required to confirm this explanation. 

This dramatic increase in the [OIII] luminosity of quasar hosts compared to $z\sim2$, and therefore the inferred equivalent width of the line,
is similar to what has been inferred for star-forming galaxies \citep{Endsley} who present evidence that at redshifts$\sim7$,
50\% of star-forming galaxies have rest-frame [OIII] equivalent widths greater than 500\AA. Since it is challenging to obtain low equivalent width H$\beta$ and strong [OIII] emission purely from photoionization models \citep{bas05}, the most likely explanation is due to strong outflows powering radiative shocks.

\subsection{An estimate of magnetic field strengths}

Nebular line ratios in narrow-line regions of 
AGN are typically explained by strong radiative shocks resulting from outflows into the surrounding ISM \citep[e.g][]{all08}.  It is generally argued that the scatter in the BPT diagram is driven by the strength and hardness of the ionizing radiation field, metallicity, gas density and shock parameters. \citet{Stern2013} show that in Type I AGN which our systems are, the [OIII]/H$\beta$ remain constant even if one separates the NLR region emission from the BLR region, while the [NII]/H$\alpha$ typically decreases. Since we do not have the kinematic information to distinguish the relative intensities of the two components, we characterize the physical properties derived from the integrated emission. This provides a lower limit to the emission line ratios of the narrow-line region.

Due to the high [OIII] to bolometric luminosity ratio of $\sim$0.1 seen in our sample, it appears that the mechanism for powering the line emission is not the same as the broad line-width quasar-powered outflows seen at $z\sim2$. Those objects, showed [OIII] to bolometric luminosity ratios of $\sim$10$^{-3}$
and had [OIII] luminosities of a few times 10$^{43}$ erg\,s$^{-1}$ . They showed [OIII] narrow to
broad line ratios of 1:2 or 1:3 at comparable bolometric luminosities. If the bulk of our observed [OIII] emission is instead from star-formation driven winds, the ionizing photons from star-formation would also contribute to the H$\alpha$ and H$\beta$ emission. Since high redshift starbursts are thought to have low [NII]/H$\alpha$ ratios and high [OIII]/H$\beta$ ratios \citep[e.g.][]{Shim2013}, subtracting a component arising from the quasar itself would shift our measured ratios by at most 0.2 dex in either axis, up and to the right (the direction as the arrow in Figure~\ref{fig:shock}-a), but this is challenging to constrain robustly - it is instead more illustrative to compare the derived properties with rare objects that have similar [OIII]/H$\beta$ line ratios in the local Universe, as we do below. 

\begin{figure*}
\centering
\includegraphics[width=6.5in]{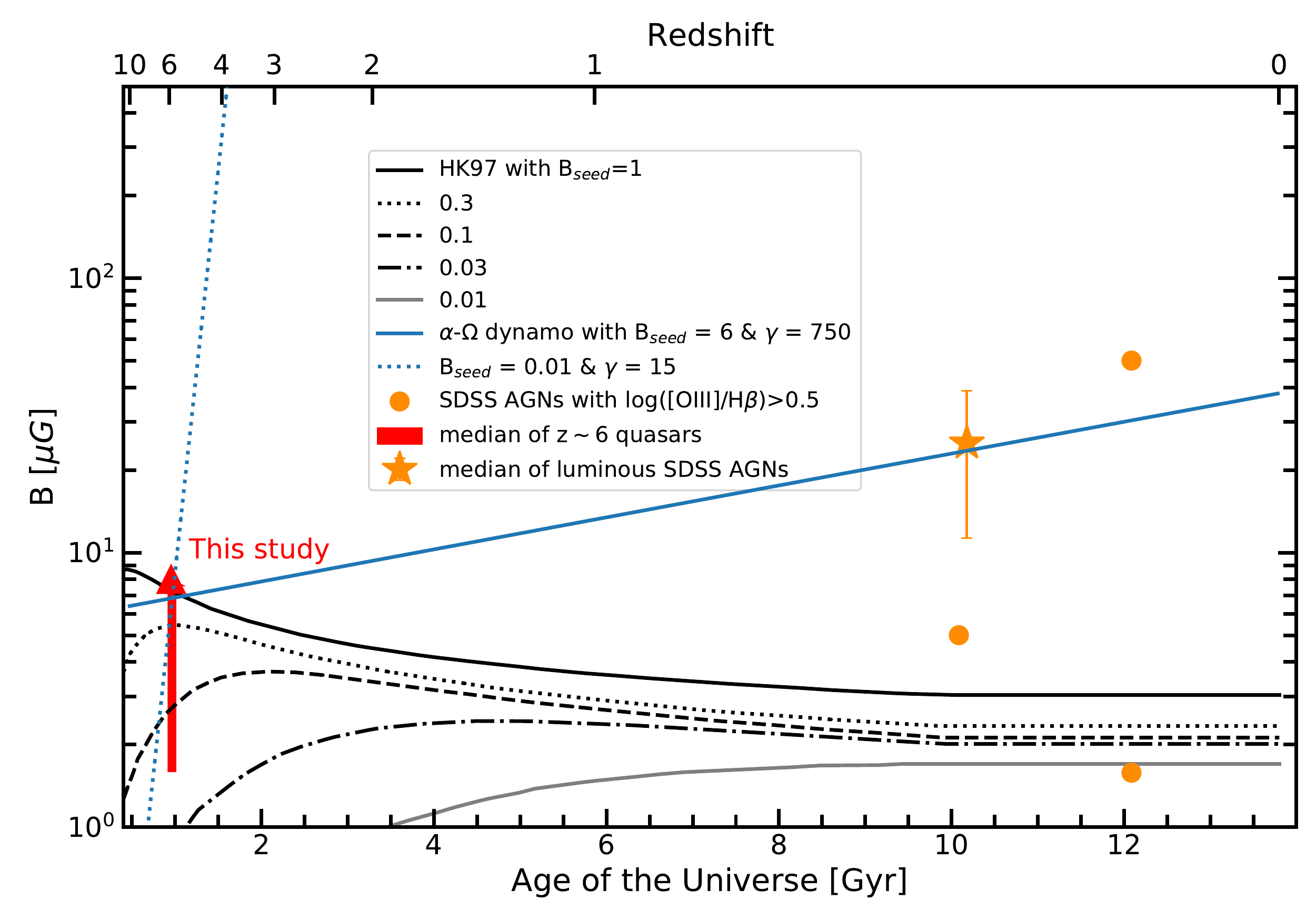}
\caption{Our constraints on the derived magnetic field in quasars at z$\sim$6 are shown as the red arrow. The median derived magnetic field for luminous local AGNs is shown as the orange star (with uncertainties). The derived magnetic field for the three luminous Sloan AGNs with log([OIII]/H$\beta$) $> 0.5$ is shown as the orange circles. Overplotted are different models for the evolution of galaxy-scale seed magnetic fields. The blue dotted and blue solid line are for the alpha-$\Omega$ dynamo model with two vastly different seed fields and dynamical timescales, forced to fit our observational constraint. The black lines are the evolution of seed fields from \protect\cite{how97} with seed values of 1, 0.3, 0.1, 0.03 and 0.01 $\mu$G. The derived magnetic field strength of z$\sim$6 quasars argues for a seed field exceeding 0.1 $\mu$G. Such a high seed is unlikely to be primordial and likely arose from turbulence in the interstellar medium during the initial stages of galaxy formation \protect\citep{kul97}.}
\label{fig:BB}
\end{figure*}

We use a library of fully radiative shock models calculated with the MAPPINGS III shock and photoionization code from \cite{all08} to interpret our measures of emission line flux ratios. The library consists of grids of models with shock velocities, magnetic fields, and pre-shock density for the solar abundance model. 
The typical volume average gas density in the NLR is expected to be $n > 100cm^{-3}$ which we adopt \citep{bas05}.
Here, we use the precursor + shock set of models with $n=100cm^{-3}$ and solar metallicity since they most closely match realistic expectations for the physical conditions. Since the grids of models are sparsely distributed in a discretized grid, we do a linear interpolation to fill the gap between grids. We then find the best-matched shock velocity (V$_{s}$) and magnetic field (B) by minimizing the $\chi^{2}$ statistics between the shock models and our observed line ratios as shown in Figure~\ref{fig:shock}-a). As a result, we find B = 8 $\mu$G and $V_{s}$ = 415 $km/s$ for the median line ratios of the $z\sim 6$ quasars. We note that these are galaxy-wide field strengths averaged over the narrow-line region of the halo and do not correspond to the Gauss-strength fields in the immediate vicinity of the supermassive black hole which has been probed by the Event Horizon Telescope \citep{EHT2021}.
Note that if the gas density is higher, we find B= 316 $\mu$G and $V_{s}$ = 255 $km/s$ with $n=1000\,cm^{-3}$.

For the median of the narrow-line regions of the luminous Sloan AGNs, we derive B = 25.1 $\mu$G and $V_{s}$ = 187.5 $km/s$, with the values for the individual luminous Sloan AGNs shown in Figure~\ref{fig:shock}-b).  Comparing values with similar assumptions on gas densities, we find the typical luminous Sloan AGNs in the local Universe appear to have larger B-fields and lower shock velocities, as derived from our fits to the integrated emission line ratios. 

However, for the subsample of Sloan AGN (blue dots) having log([OIII]/H$\beta$) $>0.5$, we find shock velocities similar to those that we find in high-z quasars, while the best-fit B-field values are in the range 1.5 to 50 $\mu$G, straddling the best fit that we derive for the high-z quasars. Due to the fact that the [FeII]/[MgII] ratios in luminous quasars appear to have a narrow range across a wide range of redshifts, and simulations argue for high stellar masses and solar metallicities, we conclude that the physical parameters that we derive for this subsample of Sloan AGN is representative of the physical parameters for the high-z quasars and conclude a lower limit to the magnetic field strength to be 1.5 $\mu$G with the best estimate being 8 $\mu$G.
Our results argue for high shock velocities and strong magnetic fields in the primordial Universe, similar to that seen in local luminous AGNs.

\subsection{Evolution of cosmic magnetic fields}

Our result places the first constraints on the amplitude of magnetic fields in massive halos that harbor $>10^{10}$\,M$_{\sun}$ galaxies and $\sim10^{9}$\,M$_{\sun}$ super-massive black holes within 1 billion years of the Big Bang. We interpret these results in the context of the expected evolution of seed fields into galactic magnetic fields.

These halos are likely only about 1 kpc in radius, indicating that they have only undergone about $\sim$200 dynamical timescales over their lifetimes by $z\sim6$. If primordial seed fields from the inflation era were present, they are thought to be of order 10$^{-17}-10^{-20}$ G and would only be amplified by a small factor within 1 Gyr. Such an evolution would be unable to account for the microGauss fields that we find in these halos.

In contrast, if there were turbulence in the interstellar medium (ISM) during structure formation, this can result in the formation of current eddies which produce a protogalactic seed magnetic field with amplitudes of order 0.01-1 $\mu$G \citep{Biermann1951, Ruzmaikin1988, kul97}. This field  is large enough, that it can easily get amplified to our derived seed field strengths, as baryons collapse into dark matter halos as has been estimated by \cite{how97} and is shown in Figure~\ref{fig:BB}. As the seed field strengthens through structure formation, dissipative processes come into play which cause the amplified field strength to decline.
Our derived field strength implies that the seed field had to be as high as 0.1 $\mu$G to be amplified to $>$1 $\mu$G by $z\sim6$.

A weaker seed field ($\sim$0.01 $\mu$G) with a higher amplification factor by the alpha-$\Omega$ dynamo wildly overpredicts the derived fields for low-z quasars as shown in Figure~\ref{fig:BB}. Such a seed evolves as \citep{Ruzmaikin1988}:
\begin{equation}
B = B_{\rm{seed}}\times\exp\left( \frac{t}{\gamma\tau_{\rm rot}}\right )
\end{equation}
where B$_{\rm{seed}}$ is the seed field, $t$ is the age of the Universe, $\gamma$ is the number of dynamical timescales while $\tau_{\rm rot}$ is the typical winding timescale expected to be $\sim10^{7}$ yrs.
A stronger seed field ($\sim$6\,$\mu$G) with a longer dynamical time
would be consistent with both measurements. However, it is challenging to account for such a slow dynamical timescale since even a lower mass dark matter
halo like our Galaxy would have undergone at least that many rotations over the central kpc. The dark matter halo of a quasar, which is at least an order of magnitude more massive would likely have far exceeded the 750 dynamical times that a higher seed field would require. 
\vspace{-0.3cm}
\section{Conclusion}

We have leveraged high quality multi-wavelength photometry to constrain the spectral energy distribution of a sample of quasars at $z\sim6$, within 1 billion years of the Big Bang.
We find that the photometry in certain bands sampled by the {\it Spitzer} and WISE spacecraft, 
is skewed high due to the entrance of strong, redshifted nebular emission lines. We use the excess in these bands to derive emission line fluxes in H$\alpha$, H$\beta$ and [OIII]. We leverage
other data to constrain the [NII] emission line flux and derive optical line ratios. We find that although the H$\alpha$ and H$\beta$ equivalent widths are consistent with other samples of quasars, the [OIII] luminosity is two orders of magnitude higher. Such high luminosities are likely powered by star-formation
that drives shocked outflows in the quasar host.
We fit the line ratios with radiative shock models and derive magnetic field strengths for our sample of quasars. We derive shock velocities of 400 km/s and integrated magnetic field strengths of $\sim$8$\,\mu$G similar to the most luminous AGNs in the local Universe.
The high derived field strength at early cosmic times argues in favor of a scenario where turbulence during structure formation seeded a field of $\sim$0.1$\mu$G which then
gets amplified as the galaxy builds its stellar mass. It is very challenging for seed fields left over as relics from the inflation era to be amplified to $\mu$G intensity by z$\sim$6.
Future, high quality mid-infrared spectroscopy, such as with JWST, will be crucial to measure the kinematics of these lines to distinguish the contribution of the BLR, and NLR and to accurately constrain
the metallicity and gas density, in addition to the magnetic field strength.

\section*{Acknowledgements}
This work is partly funded by NASA/{\it Euclid} grant 1484822 and is
based on observations taken by the {\it Spitzer}, {\it WISE} and a number of ground-based telescopes such as Pan-STARRS, UKIRT and VISTA. This work is also supported by the National Research Foundation of Korea (NRF) grant funded by the Korea government(MSIT) No. 2022R1C1C100869511. 

\section*{Data Availability}
The data underlying this article are available in the article.

\clearpage
\appendix

\section{List of quasars at $z>5$}

In Appendix, we list the photometry and derived line fluxes of the eight $z>5$ quasars used in our analysis. These have signal to noise ratio in the H$\alpha$ and [OIII] line $> 3$. In addition, we list SDSSJ0231-0728 at $z=5.37$ which is used to verify the adopted H$\alpha$/H$\beta$ ratio in Section 3.1. 
Table~\ref{table:first}, \ref{table:second}, \ref{table:third} include the J2000 coordinate, spectroscopic redshift, near-IR and mid-IR photometry where available for each object, and derived H$\alpha$ and [OIII] line fluxes in this study. 

\setcounter{table}{0}
\renewcommand{\thetable}{A\arabic{table}}
\begin{table*}
\caption{Quasars at $z>5$: near-IR photometry with J2000 coordinates  and spectroscopic redshifts.\label{table:first}}
\centering
\begin{tabular}{cccccccc}
\hline \hline
Name & RA [deg] & Dec [deg] & Redshift & y\textsuperscript{a} [AB] & J [AB]\textsuperscript{b} & H [AB]\textsuperscript{b} & K[AB]\textsuperscript{b}  \\
\hline
SDSSJ0005-0006 & 1.468 & -0.116 & 5.85 & 21.08$\pm$0.15 & 20.73$\pm$ 0.18 & 20.05$\pm$0.08 & 20.49$\pm$0.14 \\
SDSSJ0818+1722 & 124.614 &  17.381 &  6.02 & 19.30$\pm$ 0.03 &  19.09$\pm$ 0.06 & -- & -- \\
SDSSJ0836+0054& 129.183 & 0.915 & 5.81 & 19.02$\pm$0.02 &  18.65$\pm$ 0.003 & 18.36$\pm$ 0.009 & 18.11$\pm$ 0.009 \\
SDSSJ1048+4637& 162.188 & 46.622 & 6.23 & 19.61$\pm$ 0.06 & 19.10$\pm$ 0.04 & 18.92$\pm$ 0.06 & 18.94$\pm$ 0.07  \\
SDSSJ1337+4155 & 204.370 &  41.928 &5.03 & 19.35$\pm$0.03 & 19.50$\pm$ 0.09 & -- & -- \\
SDSSJ1427+3522 & 216.874 &  35.369 &  5.53  & 21.48$\pm$ 0.15 &  20.72$\pm$ 0.32 & -- & --  \\
FIRSTJ1427+3312 & 216.911 & 33.212 &  6.12 &  21.13$\pm$ 0.04 & 20.71$\pm$ 0.30 & -- & 19.93$\pm$ 0.02  \\
SDSSJ1602+4228 & 240.725 &  42.474 & 6.09 &19.86$\pm$ 0.05& 19.53$\pm$ 0.06 &  19.15$\pm$ 0.06 &  18.95$\pm$ 0.06  \\
SDSSJ0231-0728\textsuperscript{c}& 37.907 & -7.482 & 5.37 & 19.17$\pm$ 0.027 & 19.21$\pm$ 0.03 &  --  & 18.77$\pm$ 0.05  \\
\hline
\multicolumn{8}{l}{\textsuperscript{a}\footnotesize{Pan-STARRS1 DR2}} \\
\multicolumn{8}{l}{\textsuperscript{b}\footnotesize{WFCAM/UKIRT and/or VIRCAM/VISTA}}\\
\multicolumn{8}{l}{\textsuperscript{c}\footnotesize{This quasar is only used to derive H$\alpha$/H$\beta$ ratio in Section 3.1.}}\\
\bigskip
\end{tabular}

\caption{Quasars at $z>5$: \textit{Spitzer} and \textit{WISE} mid-IR photometry\textsuperscript{d} \label{table:second}}
\begin{tabular}{ccccccc}
\hline \hline
Name & IRAC 1[$\mu Jy$] & IRAC 2 [$\mu Jy$] & IRAC 3[$\mu Jy$] & IRAC 4[$\mu Jy$] & WISE 3.4$\mu m$ [$\mu Jy$] & WISE 4.6 $\mu m$ [$\mu Jy$]  \\
\hline
SDSSJ0005-0006 & 34.1$\pm$0.49 & 42.85$\pm$0.135 & 34.45$\pm$4.29& -- & 36.37$\pm$5.36 & -- \\
SDSSJ0818+1722 & 160.6$\pm$0.54 &191.2 $\pm$0.87 &  148.2 $\pm$2.83 & 191.3$\pm$ 4.93 &-- & -- \\
SDSSJ0836+0054& 264.3$\pm$0.10 & 341.2 $\pm$0.44 & 276.9 $\pm$ 0.33 &  315.5 $\pm$ 2.76 & 228.7 $\pm$8.42 & 287.1$\pm$ 13.75 \\
SDSSJ1048+4637&100.4 $\pm$0.26 & 123.3 $\pm$ 0.39 & 85.48$\pm$1.41 & 125.1 $\pm$2.32 & 87.26 $\pm$4.74 &  76.07 $\pm$8.76 \\
SDSSJ1337+4155 & 76.81$\pm$0.23& 67.11$\pm$0.35 & 65.99$\pm$1.05 & 101.8 $\pm$2.11 & 71.12 $\pm$4.78 & 38.83 $\pm$8.80 \\
SDSSJ1427+3522 &32.66$\pm$0.24 &  44.04 $\pm$0.42 & 29.02$\pm$1.73 & 61.19 $\pm$2.72 & 33.51$\pm$3.61 & 39.19$\pm$7.87 \\
FIRSTJ1427+3312 & 56.38$\pm$0.45 & 69.77 $\pm$ 0.57 & 71.38 $\pm$2.15 & 55.05$\pm$ 3.51 &  56.7 $\pm$ 3.97 & 71.65$\pm$7.92 \\
SDSSJ1602+4228 & 24.4$\pm$ 0.28& 146.1$\pm$ 0.39 & 126.7$\pm$1.43& 155.1 $\pm$2.22 & 114.8 $\pm$ 4.76 & 149.8 $\pm$8.28 \\
SDSSJ0231-0728\textsuperscript{c} & 126.50$\pm$0.35 & 173.70$\pm$0.44 & 137.2 $\pm$0.29 & 150.30 $\pm$2.36 & 123.50$\pm$ 5.69 &169.50$\pm$ 10.93 \\
\hline
\multicolumn{7}{l}{\textsuperscript{d}\footnotesize{SEIP (Spitzer Enhanced Imaging Products)}} \\
\end{tabular}

\caption{Quasars at $z>5$: derived line flux estimates \label{table:third}}
\begin{tabular}{ccccc}
\hline \hline
Name & H$\alpha$ [erg/s/cm$^{2}$] & uncertainty of H$\alpha$  & [OIII] & uncertainty of [OIII]  \\
\hline
SDSSJ0005-0006 & $1. 06\times10^{-15}$ & $2.68\times10^{-16}$ & $1.62\times10^{-15}$ & $4.56\times 10^{-16}$\\
SDSSJ0818+1722 & $4.31\times10^{-15}$ & $8.00\times10^{-16}$ & $6.14\times10^{-15}$ & $1.78\times 10^{-15}$ \\
SDSSJ0836+0054& 7.41$\times10^{-15}$ & $1.04\times10^{-15}$ & $1.18\times10^{-14}$ & $2.40\times 10^{-15}$ \\
SDSSJ1048+4637& $1.91\times10^{-15}$ & $3.93\times10^{-16}$ & $3.41\times10^{-15}$ & $7.85\times 10^{-16}$ \\
SDSSJ1337+4155 &  $4.40\times10^{-15}$ & $1.27\times10^{-15}$ & $9.11\times10^{-15}$ & $2.70\times 10^{-15}$\\
SDSSJ1427+3522 & $1.33\times10^{-15}$ & $2.00\times10^{-16}$ & $2.79\times10^{-15}$ & $4.42\times 10^{-16}$ \\
FIRSTJ1427+3312 & $2.21\times10^{-15}$ & $2.51\times10^{-16}$ & $2.85\times10^{-15}$ & $4.90\times 10^{-16}$ \\
SDSSJ1602+4228 & $ 3.44\times10^{-15}$ & $6.39\times10^{-16}$ & $4.66\times10^{-15}$ & $1.35\times 10^{-15}$ \\
SDSSJ0231-0728\textsuperscript{c} & $4.67\times10^{-15}$ & $1.24\times10^{-15}$ & $7.25\times10^{-15}$ & $2.67\times 10^{-15}$ \\
\hline
\end{tabular}

\end{table*}

\section{Absence of Systematics in the IRAC Photometry}

\begin{figure}
\centering
\includegraphics[width=\columnwidth]{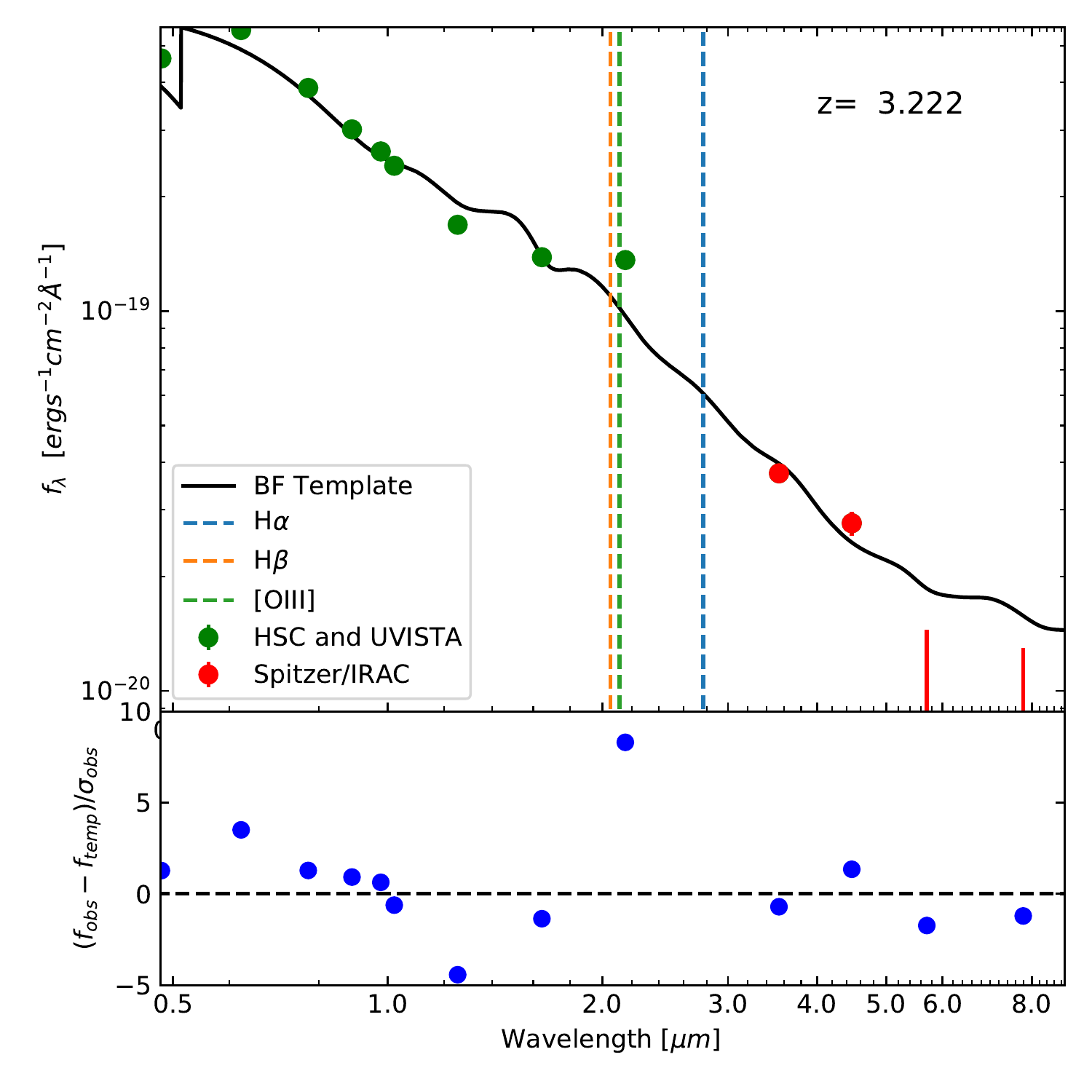}
\caption{Best-fit spectral energy distribution of a $z = 3.222$ quasar in the COSMOS field. As before, the points are the photometric measurements while the dark line is the best-fit SED. The lower panel shows the residuals between the best fit and the photometry, scaled by the photometric noise. As can be seen, the residuals in the
IRAC bands at 3.6 and 4.5$\mu$m are consistent with zero while the UVISTA/$Ks-$band shows an excess due to the entry of $H{\beta}$ and [OIII].}
\label{fig:nosystematics}
\end{figure}

The robustness of the derived line flux using our methodology requires the photometry to be free of systematics. Although we have eliminated sources which may be confused, there is the possibility that the photometry is biased in some way. In order to assess if there is a systematic bias to the IRAC photometry, we identified two $z\sim3.2$ quasars in the COSMOS field where the emission lines are outside the IRAC bands. These are X-ray detected and with spectroscopically measured redshifts. We compiled the multi-wavelength photometry for these sources from COSMOS2020 \citep{Weaver2022}. The photometric measurements include Subaru/HSC $g, r, i, z, y$, UltraVISTA $Y, J, H, Ks$, IRAC 1,2,3,4 channels where available (See Figure \ref{fig:nosystematics}). For fitting the continuum spectral energy distribution, we adopt the same method and use the same 23 AGN-dominated continuum templates described in Section 3.1.  To find the best-fit continuum template, the Ks band which might be affected by H$\alpha$, H$\beta$ and [OIII] is not included during the fitting. We find that the residuals in the IRAC bands are indeed consistent with zero while the $Ks$ band shows an excess due to the entrance of $H{\beta}$ and [OIII] into the band. If we assume the EW(H$\beta$) is consistent with redshift, we find this excess corresponds to a [OIII] line luminosity of $\sim$10$^{43}$\,erg\,s$^{-1}$, similar to that observed spectroscopically for other quasars at $z\sim2.5$ \citep{Zakamska2016b}. This proves that our methodology does not result in anomalously high line fluxes and is not prone to photometric systematic effects. The boosted flux in the broadband photometry is indeed produced by strong emission lines. 

\end{document}